\documentclass[prd,aps,floatfix,onecolumn,superscriptaddress,preprintnumbers,preprint]{revtex4-1}

\pdfoutput=1
\usepackage{mathrsfs}
\usepackage{amsfonts}
\usepackage{amsmath,amssymb,bm,bbm}
\usepackage[colorlinks=true,linkcolor=blue,citecolor=blue,urlcolor=blue]{hyperref}
\usepackage{epsfig}
\usepackage{graphicx}
\usepackage{slashed}
\usepackage{booktabs}
\usepackage{tabu}
\usepackage[dvipsnames]{xcolor}
\usepackage[normalem]{ulem}
\usepackage{tikz}
\usepackage{soul}
\usetikzlibrary{shapes,arrows}
\usetikzlibrary{positioning}
\usepackage{ulem}
\usepackage{hhline}
\usepackage{sidecap}
\sidecaptionvpos{figure}{c}
\sidecaptionvpos{table}{c}

\newcommand{\dd}[1]{\mathrm{d}#1}
\newcommand{\xb}{x_B}
\newcommand{\hel}{{^3 {\rm He}}}
\newcommand{\tri}{{^3 {\rm H}}}

\newcommand{\chired}{{$\chi^2_{\rm red}$~}}
\newcommand{\MAR}{{\footnotesize MARATHON~}}

\begin{document}

\title{\mbox{\hspace*{-0.25cm}Isospin dependence of nuclear EMC effect from global QCD analysis}}

\author{C.~Cocuzza}
\affiliation{\mbox{Department of Physics, William \& Mary, Williamsburg, Virginia 23187, USA}}
\author{T.~J.~Hague}
\affiliation{Jefferson Lab, Newport News, Virginia 23606, USA}
\author{W.~Melnitchouk}
\affiliation{Jefferson Lab, Newport News, Virginia 23606, USA}
\author{N.~Sato}
\affiliation{Jefferson Lab, Newport News, Virginia 23606, USA}
\author{A. W. Thomas}
\affiliation{\mbox{CSSM and CDMPP, Department of Physics, Adelaide University, SA 5005, Australia} \\
\vspace*{0.2cm}
{\bf JAM Collaboration \\ {\footnotesize \ (PDF Analysis Group)}
\vspace*{0.2cm} }}

\begin{abstract}
We perform a new global QCD analysis of unpolarized parton distribution functions (PDFs) in the nucleon from proton, deuteron and $A=3$ data, including recent measurements of $^3$He/$D$ and $^3$H/$D$ cross section ratios from the \MAR experiment at Jefferson Lab.
Simultaneously inferring the PDFs and nucleon off-shell corrections allows both to be determined consistently, without theoretical assumptions about the isospin dependence of nuclear effects.
The analysis provides strong evidence for the need of nucleon off-shell corrections to describe the $A=3$ data, with large isoscalar and a suggestion of nonzero isovector contributions in $A \leq 3$ nuclei. 
We find that the extracted EMC ratios of nuclear to nucleon structure functions for $A=2$ and 3 differ from those naively extrapolated from heavy nuclei down to low $A$.
\end{abstract}

\date{\today}
\preprint{JLAB-THY-26-4603, ADP-26-03/T1300}
\maketitle

\section{Introduction}
\label{s.intro}


Understanding the structure of nuclei directly in terms of the fundamental quark and gluon degrees of freedom of QCD theory remains one of the ultimate objectives of nuclear physics.
The first evidence for a modification of the parton structure of protons and neutrons inside nuclei was seen over 4 decades ago with the famous ``nuclear EMC effect''~\cite{EuropeanMuon:1983wih, Gomez:1993ri}, which revealed nontrivial deviations of the ratio of nuclear to deuteron structure functions from unity.
While a universally accepted quantitative explanation has remained elusive, some approaches have related the observed structure function modifications to the strong scalar and vector mean fields in nuclei~\cite{Thomas:1989vt, Saito:1992rm, Mineo:2003vc, Cloet:2006bq, Smith:2002ci, Guichon:2018uew}, while others have attributed the medium modification of the nucleon structure to short-range correlations~\cite{CLAS:2019vsb, Wang:2020uhj, Segarra:2020plg, Xing:2023uhj, Kim:2024wne} (see Refs.~\cite{Geesaman:1995yd, Norton:2003cb,Thomas:2018kcx} for reviews of the different approaches).

Regardless of the dynamics responsible for the medium modifications, generally the nuclear structure functions can be represented as a sum of free-nucleon structure functions, convoluted with nucleon smearing functions that take into account nuclear binding and Fermi motion, and contributions that depend on modifications of the parton structure of the bound nucleons in the nucleus~\cite{Tropiano:2018quk}.
The latter can be formally represented in terms of nucleon off-shell corrections~\cite{Dunne:1985cn, Akulinichev:1985ij, Bickerstaff:1989ch, Melnitchouk:1993nk, Melnitchouk:1994rv, Kulagin:1994cj, Kulagin:1994fz, Kulagin:2004ie}, which can be either estimated in quark models or parametrized and extracted from experiment within a given theoretical framework.

While the general features of the nuclear EMC effect for heavy nuclei can be accommodated within more than one approach, it has been difficult to obtain unique experimental signatures that could potentially distinguish between different models.
On the other hand, it has been noted that predictions for more specific aspects of the nuclear modifications, such as their spin and flavor or isospin dependence, can differ markedly depending on the dynamical mechanism assumed responsible for the modifications~\cite{Thomas:2018kcx}.
A particularly useful testing ground for studying the isospin dependence of the nuclear modifications are the light nuclei, such as the deuteron and the tri-nucleons, $^3$He and $^3$H, for which quantitative few-body calculations are more readily accessible than for nuclei with large mass number,~$A$.

In fact, over 25 years ago Afnan {\it et al.}~\cite{Afnan:2000uh, Afnan:2003vh} (see also Refs.~\cite{Pace:2001cm, Sargsian:2001gu}) suggested that deep-inelastic scattering (DIS) from the $^3$He and $^3$H mirror nuclei could be a way to constrain the $d/u$ quark distribution function ratio in the proton, assuming that the nuclear effects in the $A=3$ nuclei mostly cancel.
The availability of the high-precision data from the \MAR experiment at Jefferson Lab, firstly on the ratio of $\hel/\tri$ cross sections \cite{JeffersonLabHallATritium:2021usd} and more recently on the individual $\hel/D$ and $\tri/D$ cross section ratios \cite{JeffersonLabHallATritium:2024las}, has allowed this assumption to be tested for the first time.
In particular, by performing a global QCD analysis of the world's high-energy scattering data, in which the nucleon parton distribution function (PDF) and off-shell parameters were fitted simultaneously, Cocuzza {\it et al.} \cite{Cocuzza:2021rfn} found the first indications for an isospin dependent nuclear EMC effect in light nuclei, suggesting an enhanced nuclear effect on the $d$-quark PDF in the bound proton.
With the measurement of the individual $\hel/D$ and $\tri/D$ ratios~\cite{JeffersonLabHallATritium:2024las}, an additional experimental constraint has become available, which could in principle provide an even more robust determination of the isospin dependence of the nuclear (off-shell) corrections.

In the analysis of the \MAR data in Ref.~\cite{JeffersonLabHallATritium:2024las}, the $\hel/D$ and $\tri/D$ cross section ratios were normalized under the assumption of the same universal shape around $x = 0.3$ for the nuclear to nucleon cross section ratios across all nuclei, from Pb down to deuterium, as in the model of Kulagin and Petti (KP)~\cite{Kulagin:2004ie, Kulagin:2010gd}.
While the binding energies per nucleon for nuclei from $^4$He to $^{208}$Pb are known to be in the range (7~--~8)~MeV, those for the light nuclei $^2$H, $\hel$ and $\tri$ are considerably smaller, $\lesssim 3$~MeV, so that extrapolating the dynamics of heavy nuclear systems smoothly to $A \leq 3$ nuclei is a strong assumption.

As we demonstrate in the present work, this assumption introduces a significant bias into the analysis, leading to rather different conclusions about the nuclear EMC effect and the extracted neutron to proton structure function ratio, which was one of the goals of the \MAR experiment.
In contrast, we avoid assumptions about the shape and magnitude of the off-shell corrections in the $A \leq 3$ nuclei and allow these to be determined from the global QCD analysis.
Consistent with the previous JAM analysis~\cite{Cocuzza:2021rfn} of global data including the $\hel/\tri$ ratio, we find strong evidence for sizable off-shell contributions to the nuclear structure functions, and a flavor dependence of the EMC ratios in $A=3$ nuclei.


In this paper we present the results of a global QCD analysis of unpolarized PDFs from the world's high-energy scattering data on protons, deuterons, and $A=3$ nuclei using the JAM Bayesian Monte Carlo framework.
In particular, we simultaneously fit the free nucleon PDF and nucleon off-shell parameters, with the latter constrained by the \MAR data on the $D/p$, $\hel/D$ and $\tri/D$ structure function ratios, along with other high-energy scattering data involving $A=2$ and $A=3$ nuclei.
We begin in Sec.~\ref{s.theory} with an overview of the theoretical framework employed in this analysis, focusing in particular on the formulation of nuclear structure functions in terms of convolutions of free nucleon structure functions and nucleon off-shell corrections, along with some details of the methodology used in the analysis.
In Sec.~\ref{s.fit} we discuss the quality of the global fit to all the data, paying particular attention to the relative normalizations of the various datasets used in the analysis.
The results for the PDFs and off-shell functions from the QCD analysis are presented in Sec.~\ref{s.QCDanalysis}, along with those for ratios of nuclear structure functions determined from the global fit.
We also contrast our results with those obtained from the analysis in Ref.~\cite{JeffersonLabHallATritium:2024las} that imposes the KP model~\cite{Kulagin:2004ie, Kulagin:2010gd} as the reference point for the \MAR data normalization.
Finally, in Sec.~\ref{s.outlook} we summarize our findings and outline future avenues for improving our understanding of the relation between PDFs in free nucleons and those bound inside nuclei.

\section{Theoretical framework}
\label{s.theory}

In this section we outline the theoretical framework used in this analysis, focusing on the essential features related to the description of nuclear structure functions in lepton-nucleus DIS.
We review the standard convolution formulas for the nuclear structure functions in the nuclear impulse approximation, and discuss in detail the nucleon off-shell contributions in terms of their isoscalar and isovector components.
We further summarize the core aspects of the Bayesian Monte Carlo methodology employed in JAM global QCD analyses, along with details of the parametrizations employed for the various parton distributions.

\subsection{Nuclear structure functions}
\label{ssec.F2A}

In the nuclear impulse (or weak nuclear binding) approximation, DIS from a nucleus~$A$ (with 4-momentum $P_A$) takes place through the scattering of an exchanged virtual boson (4-momentum $q$) incoherently from individual off-shell nucleons (4-momentum~$p$) in the nucleus~\cite{Dunne:1985cn, Akulinichev:1985ij, Bickerstaff:1989ch, Melnitchouk:1993nk, Melnitchouk:1994rv, Kulagin:1994cj, Kulagin:1994fz}.
In this case the nuclear $F_2^A$ structure function can be written in convolution form as a sum of on-shell and off-shell nucleon contributions \cite{Melnitchouk:1993nk, Melnitchouk:1994rv, Kulagin:2004ie, Kulagin:2010gd, Tropiano:2018quk, Cocuzza:2021rfn},
\begin{align}
F_2^A(\xb,Q^2)
& = \sum_N 
\Big[ f_{N/A}^{\rm (on)} \otimes F_2^N
    + f_{N/A}^{\rm (off)} \otimes \delta F_2^{N/A}
\Big](x_B,Q^2),
\label{eq.F2A}
\end{align}
where $x_B = (M_A/M)\, Q^2/(2 P_A \cdot q$) is the Bjorken scaling variable, scaled by the ratio of nuclear~($M_A$) to nucleon ($M$) masses to be in the range $x_B \in [0, M_A/M] \approx [0,A]$, and $Q^2$ is the 4-momentum squared of the exchanged boson.

The functions $f_{N/A}^{\rm (on)}$ and $f_{N/A}^{\rm (off)}$ are the on-shell and off-shell smearing functions of nucleons $N$ in nucleus $A$~\cite{Tropiano:2018quk}, and $F_2^N$ and $\delta F_2^{N/A}$ are the on-shell and off-shell nucleon structure functions, respectively.
The symbol $\otimes$ in Eq.~(\ref{eq.F2A}) represents the convolution
\begin{align}
\big[ f_{N/A} \otimes F_2 \big](x_B,Q^2) 
&\equiv \int_{x_B}^{M_A/M} \dd y\, f_{N/A}(y,\rho)\, F_2\Big(\frac{x_B}{y},Q^2\Big),
\label{eq.convA}
\end{align}
where $y = (p_0 + \rho\, p_z)/M$ is the light-cone fraction of the nuclear momentum carried by the interacting nucleon, and $p_0$ and $p_z$ are the nucleon's energy and longitudinal momentum, respectively.
The kinematical parameter $\rho^2 = {1 + 4 M^2 x_B^2/Q^2}$ takes into account finite energy effects.
Finally, we note that while Eqs.~(\ref{eq.F2A}) and (\ref{eq.convA}) are valid for all nuclei $A$, in practice we will consider only the specific cases of $A=2$ (deuteron) and $A=3$ ($^3$He and $^3$H).

In QCD collinear factorization the nucleon structure functions can be written as convolutions of PDFs with the hard scattering coefficients, $C_q$,
\begin{align}
F_2^N(x_B,Q^2)
& = x_B \sum_q e_q^2\, \big[ C_q \otimes q^+_N \big](x_B,Q^2) 
  +\ \cdots
\label{eq.F2Non}
\end{align}
where $q^+_N = q_N + \bar q_N$ is the $C$-even PDF in the nucleon, and the ellipsis includes higher order contributions from gluons, as well as power corrections associated with target mass effects and higher twist (HT) multiparton contributions.
Note that here and in the following the PDFs $q^+_N$ are functions of the momentum fraction, $x$, of the nucleon carried by the parton, which at leading order coincides with the Bjorken scaling variable, but differs at higher orders.
The convolution in this case is given by
\begin{align}
\big[ C_q \otimes q_N^+ \big](x_B,Q^2) 
&\equiv \int_{x_B}^{1} \frac{\dd x}{x} C_q(x)\, q_N^+\Big(\frac{x_B}{x},Q^2\Big).
\label{eq.convN}
\end{align}
When referring to the free nucleon in Eq.~(\ref{eq.convN}) and elsewhere, it is understood that $\xb$ refers to the corresponding free-nucleon ($A=1$) scaling variable.

Analogously, we can write the off-shell nucleon structure functions in the nucleus in factorized form in terms of the off-shell PDFs $\delta q_{N/A}$ as
\begin{align}
\delta F_2^{N/A}(x_B,Q^2)
& = x_B \sum_q e_q^2\, \big[ C_q \otimes \delta q_{N/A} \big](x_B,Q^2) 
  + \cdots
\label{eq.F2Noff}
\end{align}
Similar expressions exist for the longitudinal $F_L$, transverse $F_1$ and parity-odd $F_3$ structure functions~\cite{Kulagin:2004ie, Tropiano:2018quk}.
The nuclear convolution expression in Eq.~(\ref{eq.F2A}) is obtained from the general virtual photon--off-shell nucleon scattering amplitude by Taylor expanding the bound nucleon structure function around the on-shell limit, in terms of the nucleon virtuality $v(p^2) \equiv (p^2 - M^2)/M^2$, where $p^2$ is the bound nucleon's 4-momentum squared.
%
%
Our analysis will focus on the large-$x_B$ region, where the structure functions are dominated by contributions from $u$ and $d$ quarks, and sea quark and gluon contributions to the off-shell functions can be effectively neglected.

In the leading twist approximation the nuclear structure function in Eq.~(\ref{eq.F2A}) can also be written in terms of nuclear PDFs,
\begin{align}
F_2^A(x_B,Q^2)
& = x_B \sum_q e_q^2\, \big[ C_q \otimes q^+_A \big](x_B,Q^2).
\label{eq.F2A_LT}
\end{align}
In the limit of exact charge symmetry, the $u_A$ and $d_A$ PDFs in the $A=2$ and 3 nuclei are related to each other by
\begin{eqnarray}
u_D &=& d_D,
\end{eqnarray}
and
\begin{subequations}
\begin{eqnarray}
u_{\,\hel} &=& d_{\,\tri}, \\
d_{\,\hel} &=& u_{\,\tri},
\end{eqnarray}
\end{subequations}
respectively.
In this case, the PDFs of a given flavor in different nuclei (for example, $u_D$, $u_{\,\hel}$ or $u_{\,\tri}$) are not related and are parametrized independently.
Since in this work we wish to relate the free nucleon PDFs to those in bound nucleons, we do not work with nuclear PDFs directly in Eq.~(\ref{eq.F2A_LT}), but use Eqs.~(\ref{eq.F2A})--(\ref{eq.F2Noff}) as the basis for the global analysis.

At finite $Q^2$ values the structure functions also receive contributions from target mass corrections~\cite{Brady:2011uy, Schienbein:2007gr}, for which we use the collinear factorization framework~\cite{Aivazis:1993pi, Moffat:2019qll} in order to consistently describe all of the high energy scattering data and not just inclusive DIS.
In addition, we allow for HT power corrections to the structure functions, for which we use an additive parametrization, 
\begin{align}
F_2^N(x_B,Q^2)\ &=\ F_2^{N\,({\rm TMC})}(x_B,Q^2) 
+ \frac{H^N(x_B)}{Q^2} 
+ {\cal O}\bigg(\frac{1}{Q^4}\bigg),
\label{eq.ht}
\end{align}
where $F_2^{N({\rm TMC})}$ is the leading-twist structure function with TMCs included, and $H^N(x_B)$ is the $x_B$-dependent HT function which will be parametrized in the numerical analysis.

In the literature, a multiplicative ansatz is sometimes used, in which the HT correction is multiplied by the leading twist contribution $F_2^{N\,({\rm TMC})}$.
The advantage of an additive parametrization is that it allows a cleaner separation of the $1/Q^2$ power corrections in the HTs from the $\ln Q^2$ corrections in the leading twist structure functions, whereas a multiplicate form inevitably mixes the two behaviors.
To compare with previous analyses, however, we also consider the multiplicative-like parameterization by modifying the HT function $H^N(x_B) \to H^N(x_B,Q^2) = H^N(x_B) \ln Q^2$.
In this multiplicative parametrization, the evolution of $H^N$ is determined by the leading twist evolution of $F_2^{N\,({\rm TMC})}$, for which the dependence on $Q^2$ scales with $\ln Q^2$, and which therefore allows one to mimic a multiplicative form.

We note also that while the HT contribution enters at the free-nucleon level, when applied to nuclei the HT part of the structure function is smeared by the nuclear light-cone momentum distribution as in Eq.~(\ref{eq.convA}) in the same manner as the leading twist component.
We do not include off-shell corrections to the HT term, however, since this doubly suppressed effect cannot be constrained by existing data.

\subsection{Nucleon off-shell corrections}
\label{ssec:offshell}

While the on-shell and off-shell smearing functions can be computed directly from known nuclear wave functions, the off-shell nucleon PDFs $\delta q_{N/A}$ are {\it a priori} unknown and need to be determined from experiment.
In the absence of charge symmetry violation and Coulomb effects, the off-shell contribution from $u$ quarks in a neutron inside a nucleus $A$ is identical to the off-shell contribution from $d$ quarks in a proton inside the corresponding mirror nucleus~$A^*$ (number of protons and neutrons interchanged), and similarly for $d$ quarks in a neutron,
\begin{align}
\delta u_{n/A} = \delta d_{p/A^*},& \qquad 
\delta d_{n/A} = \delta u_{p/A^*}.
\end{align}
For the specific case of the deuteron $D$,
\begin{align}
\delta u_{n/D} = \delta d_{p/D},& \qquad
\delta d_{n/D} = \delta u_{p/D},
\end{align}
while for the $A=3$ nuclei $\hel$ and $\tri$, one has
\begin{subequations}
\begin{align}
\delta u_{n/\hel} = \delta d_{p/\tri},& \qquad 
\delta d_{n/\hel} = \delta u_{p/\tri},
\\
\delta u_{n/\tri} = \delta d_{p/\hel},& \qquad 
\delta d_{n/\tri} = \delta u_{p/\hel}.
\end{align}
\end{subequations}

To further classify the off-shell distributions, we can decompose the functions into contributions reflecting the nature of the interactions between the remnants of the struck nucleon and the rest of the nucleus during the scattering process.
To illustrate this, it is instructive to consider the formal expression for the twist-two PDF of a nucleon at rest~\cite{Jaffe:1983hp},
\begin{equation}
q(x) \, = \, \sqrt{2} \sum_k \delta\big(x M - (p^+ \, - \, p_k^+)\big)
\big| \langle S_k(p_k) \, | \, 
\psi_+(0) \, 
| N(p) \rangle 
\big|^2,
\label{Eq.q}
\end{equation}
where the state $|N(p) \rangle$ is a nucleon in the Lorentz scalar and vector fields generated by the rest of the nucleus, and $|S_k(p_k) \rangle$ are the intermediate (spectator) states with 4-momentum~$p_k$ and mass $m_k$ left when a quark is removed by the operator $\psi_+$~\cite{Jaffe:1983hp}.
Note that since DIS at leading twist involves the imaginary part of the virtual Compton scattering amplitude, the spectator states are necessarily on-shell.
The key features of the valence quark PDFs of the free nucleons can be readily understood in terms of the contribution from the lowest mass intermediate state, consisting of two valence quarks \mbox{($k=2$)}, with mass $m_2$ and three-momentum $\bm{p}_{2}$~\cite{Close:1988br}. 
For ground state baryons, the most likely momentum is zero, $\bm{p}_{2} = 0$, so the $\delta$-function in Eq.~(\ref{Eq.q}) will produce a peak in the $q(x)$ distribution from this lowest mass intermediate state at $x = 1 - m_2/M$. 
The gluon hyperfine interaction, which lifts the mass of a pair of quarks with spin-1 by roughly 200~MeV above that for a spin-0 pair, then naturally explains, for example, why the $d/u$ PDF ratio decreases rapidly as $x \to 1$.

In a nucleus, the off-shell corrections are expected to be dominated by the shifts in $p^+$ and $p_2^+$ resulting from their interaction with the scalar and vector fields generated by the rest of the nucleus. 
In a simple effective hadronic model these interactions may be considered as generated by the exchange of isoscalar $\sigma$ and $\omega$ mesons and the isovector $\rho$ meson.
Labeling the potentials felt by each quark due to the exchange of these mesons as $V_\sigma$, $V_\omega$ and $V_\rho$, the $\delta$-function in Eq.~(\ref{Eq.q}) now shifts the peak in the PDF by an amount proportional to $(V_\sigma \, + \, V_\omega \, + \, V_\rho)/M$. 
The isoscalar interactions depend only on the number of spectator nucleons, while the isovector term will depend on the third component of their isospin; for example, repulsive for a $d$-quark in a struck proton in $\tri$.

With this picture in mind, the off-shell PDF can in general be written as a sum of isoscalar~``$(0)$'' and isovector ``$(1)$'' contributions,
\begin{eqnarray}
\delta q_{N/A} &=& \delta q_{N/A}^{(0)} + \delta q_{N/A}^{(1)},
\end{eqnarray}
characterizing the possible isospin projections resulting from the interactions between the on-shell spectator quark system after a quark has been removed from the nucleon and the residual nuclear system that does not take part in the scattering.
For the case of a proton, scattering from a $u$ quark gives a spectator quark system that is dominated by a $ud$ quark pair that has third component of isospin 0, whereas scattering from a $d$ quark leaves behind a $uu$ pair that has isospin projection $+1$, and similarly for the neutron.

For a deuteron target, the recoil nuclear system consists of an isospin-1/2 nucleon, regardless of whether the proton or neutron is probed in the scattering.
In this case the spectator systems can interact via the exchange of either isoscalar or isovector quantum numbers.
For~$\tri$, scattering from a proton leaves behind two neutrons, which have isospin projection $-1$, while scattering from a neutron involves a $pn$ state with isospin projection 0.
For $\hel$, on the other hand, scattering from a proton gives an isospin-0 $pn$ state, while scattering from a neutron will leave two protons with isospin projection $+1$ in the intermediate state.
Finally, to preserve the number of valence quarks in the bound nucleons in nuclei each of the isoscalar and isovector off-shell contributions to the off-shell functions, $\delta q_{N/A}$, must satisfy
\begin{eqnarray}
\int_0^1 \dd x~\delta q_{N/A}^{(0)}(x)\ 
 =\ 0\ 
 =\ \int_0^1 \dd x~\delta q_{N/A}^{(1)}(x).
\label{eq.offsumrule}
\end{eqnarray}

In the previous JAM analysis~\cite{Cocuzza:2021rfn} of the earlier \MAR data~\cite{JeffersonLabHallATritium:2021usd} on the $\hel/\tri$ ratio, the off-shell functions $\delta q_{N/A}$ were assumed to arise purely from the isovector contributions in the nucleus.
Here we generalize this approach by allowing both isoscalar and isovector contributions to the off-shell functions.
From the above discussion, for the isoscalar contributions to the $u$ and $d$ off-shell functions, we have
\begin{subequations}
\begin{eqnarray}
\delta u^{(0)}_{p/\hel} &=& \delta u^{(0)}_{p/\tri}\ 
                         =\ 2\, \delta u^{(0)}_{p/D}\hspace*{0.1cm}
\equiv\ \delta u_0,
\\
\delta d^{\,(0)}_{p/\hel} &=& \delta d^{\,(0)}_{p/\tri}\ 
                         =\ 2\, \delta d^{\,(0)}_{p/D}\hspace*{0.14cm}
\equiv\ \delta d_0.
\end{eqnarray}
\label{eq.delta0}%
\end{subequations}
In $\hel$, the isovector contributions will be zero since the spectator system for scattering from the proton in a $\hel$ nucleus is an isospin-0 $pn$ pair; for $\tri$, on the other hand, isovector exchange is allowed,
\begin{subequations}
\begin{eqnarray}
\delta u^{(1)}_{p/\hel} &=& 0,
\\
\delta d^{\,(1)}_{p/\hel} &=& 0,
\\
\delta u^{(1)}_{p/\tri} &=& 2\, \delta u^{(1)}_{p/D}\ \equiv\ \delta u_1,
\\
\delta d^{\,(1)}_{p/\tri} &=& 2\, \delta d^{\,(1)}_{p/D}\ \equiv\ \delta d_1.
\end{eqnarray}
\label{eq.delta1}%
\end{subequations}
Using Eqs.~(\ref{eq.delta0}) and (\ref{eq.delta1}), we can therefore write the set of off-shell $u$ and $d$ PDFs in the proton in the $A=2$ and $A=3$ nuclei as 
\begin{subequations}
\begin{align}
\delta u_{p/D}\   &=\ \frac{1}{2} (\delta u_0 + \delta u_1),
\qquad
\delta u_{p/\hel}\ =\ \delta u_0,
\qquad
\delta u_{p/\tri}\ =\ \delta u_0 +  \delta u_1,
\\
\delta d_{p/D}\   &=\ \frac{1}{2} (\delta d_0 + \delta d_1),
\qquad\,
\delta d_{p/\hel}\ =\ \delta d_0,
\qquad\,
\delta d_{p/\tri}\ =\ \delta d_0 +  \delta d_1,
\end{align}
\end{subequations}
%
%
so that all of the off-shell functions can be expressed in terms of the four basis off-shell functions $\delta u_0$, $\delta d_0$, $\delta u_1$, and $\delta d_1$.
However, since there are only three unique experimental constraints from the inclusive nuclear DIS data (on $D$, $\hel$, and $\tri$ nuclei), the number of functions that can be determined from data needs to be reduced to at most three.
%
%
Naturally, the reduction of the number of functions to be determined should be made in the most physically motivated way possible.
Setting $\delta u_1$ equal to $\delta d_1$, for instance, would contradict expectations from mean field calculations of nuclear forces~\cite{Thomas:1989vt, Saito:1992rm, Mineo:2003vc, Cloet:2006bq, Guichon:2018uew}, so it is more reasonable to set the isoscalar functions equal, $\delta u_0 = \delta d_0 \equiv \delta q_0$.
This allows us to obtain the most faithful representation of the uncertainties possible on quantities that are sensitive to off-shell corrections, such as nuclear EMC ratios, while taking into account the limits of the experimental data and keeping the model physically motivated.

We have also tested two other scenarios with isovector off-shell corrections switched off, $\delta u_1 = \delta d_1 = 0$, namely, one with $\delta u_0 \neq \delta d_0$ and one with $\delta u_0 = \delta d_0$.
For the former, we find that the experimental data cannot distinguish between the individual $\delta u_0$ and $\delta d_0$ flavors, and these end up being highly correlated.
This further motivates the choice of setting the isoscalar off-shell functions equal and fitting the three independent functions $\delta q_0$, $\delta u_1$, and $\delta d_1$.
In practice, we find that the extracted isoscalar function $\delta q_0$ is very similar to the isoscalar off-shell contribution extracted when isovector effects are set to zero, which demonstrates the stability of the extracted isoscalar component.

\subsection{Methodology}
\label{s.methodology}

The methodology employed in this analysis is based on Bayesian inference using Monte Carlo techniques developed in previous JAM QCD global analyses \cite{Sato:2016tuz, Sato:2016wqj, Ethier:2017zbq, Sato:2019yez, Moffat:2021dji, Cocuzza:2021rfn, Cocuzza:2021cbi, Anderson:2024evk}.
In contrast to attempts to extract information on PDFs or nuclear effects from a single experiment~\cite{JeffersonLabHallATritium:2021usd, JeffersonLabHallATritium:2024las}, which inevitably requires model-dependent inputs and assumptions, our global analysis allows both the nucleon PDFs and nuclear effects to be determined simultaneously with reduced theoretical bias.

The Bayesian analysis consists of sampling the posterior distribution $\mathcal{P}$ for a set of parameters $\bm{a}$ given by
\begin{align}
\mathcal{P}(\bm{a}|{\rm data})
\propto \mathcal{L}(\bm{a},{\rm data})\, \pi(\bm{a}),
\end{align}
with a likelihood function of Gaussian form,
\begin{align}
\mathcal{L}(\bm{a},{\rm data}) 
= \exp\!\Big(\! \!-\!\frac12 \chi^2(\bm{a},{\rm data})\! \Big),
\label{eq.likelihood}
\end{align}
and a prior function $\pi(\bm{a})$.
The $\chi^2$ function in Eq.~(\ref{eq.likelihood}) is defined for each replica as
\begin{align}
\label{eq.chi2}
\chi^2(\bm{a}) &= \sum_{e,i} 
\bigg( 
\frac{d_{e,i} - \sum_k r_{e,k} \beta_{e,i}^k - T_{e,i}(\bm{a})/N_e}{\alpha_{e,i}} \bigg)^{\!2},
\end{align}
where $d_{e,i}$ is the data point $i$ from experimental dataset $e$, and $T_{e,i}(\bm{a})$ is the corresponding theoretical value.
All uncorrelated uncertainties are added in quadrature and labeled by $\alpha_{e,i}$, while $\beta_{e,i}^k$ represents the $k$-th source of point-to-point correlated systematic uncertainties for the $i$-th data point weighted by $r_{e,k}$.
The latter are optimized per values of the parameters~$\bm{a}$ via $\partial \chi^2/\partial r_{e,k}=-2r_{e,k}$, which can be solved in closed form.
We include normalization parameters $N_e$ for each dataset $e$ as part of the posterior distribution per dataset.

The prior $\pi(\bm{a})$ for each replica is given by
\begin{align}
\pi(\bm{a}) &= 
{\bigg[ \prod_\ell \Theta 
\big( 
    (a_\ell-a_\ell^{\rm min}) (a_\ell^{\rm max}-a_\ell) 
\big) 
\bigg] 
\bigg[ \prod_e \prod_k 
    \exp{\left(\! -\frac{1}{2} r_{e,k}^2 \right)}
\bigg]}
{\bigg[ \prod_e 
\exp{ 
    \bigg(\!\!-\frac{1}{2} 
    \Big( \frac{1-N_e}{\delta N_e} 
    \Big)^{\!2} 
    \bigg)} 
\bigg]},
\label{e.prior}
\end{align}
where $\prod_\ell$ is a product over the parameters.
The step function $\Theta$ forces the parameters to be within the chosen range between $a_\ell^{\rm min}$ and $a_\ell^{\rm max}$.
A Gaussian penalty controlled by the experimentally quoted normalization uncertainties $\delta N_e$ is applied when fitting the normalizations $N_e$.
Similarly, a penalty proportional to $r_{e,k}^2$ is applied when the experiment $e$ has correlated sources of uncertainty.

As in most previous unpolarized PDF analyses by JAM and other groups, we parametrize the PDFs at the input scale $\mu_0^2 = m_c^2$ using a generic template function of the form
\begin{align}
f(x,\mu_0^2) 
= \frac{N}{\cal M}\, x^{\alpha}(1-x)^{\beta}(1+\gamma \sqrt{x} + \eta x),
\label{eq.template}
\end{align}
where $\bm{a} = \{ N, \alpha, \beta, \gamma, \eta \}$ is the set of parameters to be inferred.
The normalization constant
    ${\cal M} = {\rm B}[\alpha+2,\beta+1]
              + \gamma {\rm B}[\alpha+\frac52,\beta+1]
              + \eta {\rm B}[\alpha+3,\beta+1]$,
where $\rm B$ is the Euler beta function, normalizes the function to the second moment, and allows us to avoid large numerical values for the $\alpha$ and $\beta$ parameters.

Our parameterization of the onshell unpolarized PDFs is similar to that of the recent JAM analysis of Ref.~\cite{Anderson:2024evk}.
To characterize the nucleon valence region and discriminate it from the sea components, we parametrize the light-quark and strange PDFs according to
\begin{align}
u      &= u_v + \bar{u},    \hspace{0.95cm}
d       = d_v + \bar{d},    \notag\\         
\bar{u}&= S_1 + \bar{u}_0,  \hspace{0.80cm}
\bar{d} = S_1 + \bar{d}_0,  
\label{eq.param}             \\
s      &= S_2 + s_0,        \hspace{0.85cm}
\bar{s} = S_2 + \bar{s}_0,  \notag
\end{align}
where the dependence on $x$ and the scale $\mu_0^2$ has been suppressed for convenience.
The input quark distributions $u_v$, $d_v$, $\bar{u}_0$, $\bar{d}_0$, $s_0$, and $\bar{s}_0$, as well as the gluon distribution $g$, are parametrized individually as in Eq.~(\ref{eq.template}). 
For the sea quark PDFs, the additional functions $S_1$ and $S_2$ are also parametrized via Eq.~(\ref{eq.template}), and are designed to allow a more singular small-$x$ behavior compared to the valence distributions by restricting the corresponding $\alpha$ parameter to more negative values.
For $u_v$, $d_v$, $s$, and $g$ the normalization parameters $N$ are fixed by the valence and momentum sum rules.
For $s$, $\bar{s}$, $S_1$, and $S_2$ we also set $\gamma = \eta = 0$.
This leads to a total of 33 free parameters for the onshell PDFs.

For the off-shell functions, we fit the isoscalar $\delta q_0$ and isovector $\delta u_1$ and $\delta d_1$ functions independently using Eq.~(\ref{eq.template}), with $\gamma = 0$ and $\eta$ fixed by the sum rule (\ref{eq.offsumrule}). 
This leads to three independent parameters for each function, and a total of nine for the entire set of off-shell functions.

For the HT corrections to the DIS structure functions, we use the additive parametrization in Eq.~(\ref{eq.ht}), in contrast to the previous JAM analysis of the \MAR data~\cite{JeffersonLabHallATritium:2021usd} where a multiplicative parametrization was used.
In practice, since our HT parametrization is sufficiently flexible, we find that the choice of additive or multiplicative ansatz does not substantially affect the final results.
However, as discussed in Sec.~\ref{ssec.F2A}, we choose the additive form in order to maintain a cleaner separation between the $1/Q^2$ power and logarithmic in $Q^2$ corrections in the HTs.
The HT corrections for protons and neutrons in the present work are parametrized  independently, using Eq.~(\ref{eq.template}) but without the factor $\mathcal{M}$.
This leads to a total of 10 independent parameters for the HT corrections.
In addition to these parameters, there are a further 22 fitted normalizations in the analysis, leading to a total of 74 free parameters in the global fit.

\section{Quality of fit and data normalization}
\label{s.fit}

The baseline data used in this analysis are very similar to those used in the previous JAM global analysis~\cite{Cocuzza:2021rfn} that included the earlier \MAR data~\cite{JeffersonLabHallATritium:2021usd}, but with a few updates. 
Specifically, the $W^2$ cut for unpolarized DIS data was increased from 3.0~GeV$^2$ to 3.5~GeV$^2$, which significantly improved the \chired for some datasets.
Inclusive $W$+charm data were also included, as in the recent JAM study in Ref.~\cite{Anderson:2024evk}.
In addition to the previous baseline, we also include the recent $D/p$ inclusive DIS data from the E12-10-002 experiment in Hall~C at Jefferson Lab~\cite{HallC:2024vvy}.

While the \MAR data on the $D/p$ and $\hel/\tri$ ratios \cite{JeffersonLabHallATritium:2021usd} were included in the earlier JAM global analysis~\cite{Cocuzza:2021rfn}, the current study includes the new data on the individual $\hel/D$ and $\tri/D$ cross section ratios~\cite{JeffersonLabHallATritium:2024las} in place of the $\hel/\tri$ measurement.
The restriction on the $W^2$ range removes the highest-$x_B$ point in both of the $\hel/D$ and $\tri/D$ measurements, so that 21 of the 22 original data points for both ratios are included.
Crucially, whereas the $\hel/D$ and $\tri/D$ ratios in the \MAR analysis~\cite{JeffersonLabHallATritium:2021usd} were normalized by factors of 1.021 and 0.996, respectively, to match with the KP model~\cite{Kulagin:2010gd}, in our analysis we do not impose any model-dependent normalization on the \MAR data, but fit the overall normalizations with $\chi^2$ penalties dependent on the experimental normalization uncertainties, as detailed in Sec.~\ref{s.methodology} above.
This allows us to perform an analysis that is not biased by the assumed veracity of the KP model that was tuned to large-$A$ nuclear data~\cite{Kulagin:2010gd}.

Note that the \MAR experiment measured ratios of inclusive cross sections, without separating the transverse and longitudinal contributions.
At the 11~GeV Jefferson Lab kinematics at which the data were taken, we expect the contributions from longitudinal cross sections to be much smaller than the transverse cross sections, and the nuclear dependence of the longitudinal to transverse ratio to be weak.
Consistent with other studies of nuclear EMC ratios, in this analysis we take the \MAR cross section ratios to be the same as the structure function ratios, and refer to these interchangeably.

In Table~\ref{t.chi2} we summarize the \chired values for the new analysis, as well as for a purely on-shell analysis as a reference point.
Between these two scenarios there is almost no change for non-DIS processes, nor for HERA or the fixed target data outside of \MAR\!\!.
For the \MAR data, little change is seen for the $D/p$ ratio, while the description of the $\tri/D$ data improves.
The largest change is seen in the description of the $\hel/D$ data, which improves dramatically with inclusion of off-shell effects.
This improvement provides the strongest evidence to date for off-shell effects when describing DIS from $A \leq 3$ nuclei.

\begin{table}[b]
\caption{Reduced \chired values for the data used in this analysis, for the full fit including off-shell corrections and for the on-shell only fit. The \MAR and JLab E03-103 data are listed separately from the rest of the fixed target DIS data.}
\begin{tabular}{l r | c c }
\hhline{====}
& & \multicolumn{2}{c}{\chired}
\\
Process & ~$N_{\rm dat}$~ & ~on-shell fit~ & ~full fit~  
\\ \hline
DIS       & & & 
\\
~~~{\scriptsize MARATHON} {\footnotesize $D$}/$p$ \cite{JeffersonLabHallATritium:2021usd}      
& 7~~    & 0.68      & 0.70
\\
~~~{\scriptsize MARATHON} {\footnotesize $\hel$}/$D$ \cite{JeffersonLabHallATritium:2024las}                    
& 21~~   & 5.10      & 1.04   
\\
~~~{\scriptsize MARATHON} {\footnotesize $\tri$}/$D$ \cite{JeffersonLabHallATritium:2024las}                    
& 21~~   & 2.20      & 0.83   
\\
~~~{\footnotesize JLab E03-103} {\footnotesize $\hel$}/{\footnotesize $D$} \cite{Seely:2009gt}  
& 13~~    & 0.33      & 0.22
\\
~~~other fixed target       & ~~~~2793~~   & 0.99    & 0.99      \\
~~~{\scriptsize HERA}       & 1185~~       & 1.17    & 1.16      \\ 
Drell-Yan                   & 205~~        & 1.18    & 1.18      \\
$W$-lepton asymmetry        & 70~~         & 0.82    & 0.82      \\
$W$ charge asymmetry        & 27~~         & 1.03    & 1.03      \\
$Z$ rapidity                & 56~~         & 1.10    & 1.11      \\
jet                         & 200~~        & 1.12    & 1.10      \\ 
\hline
\textbf{Total}              & {\bf 4598}~~ & {\bf 1.07} & {\bf 1.04}    \\
\hhline{====}
\end{tabular} 
\label{t.chi2}
\end{table}

\begin{figure}[t]
\includegraphics[width=1.01\textwidth]{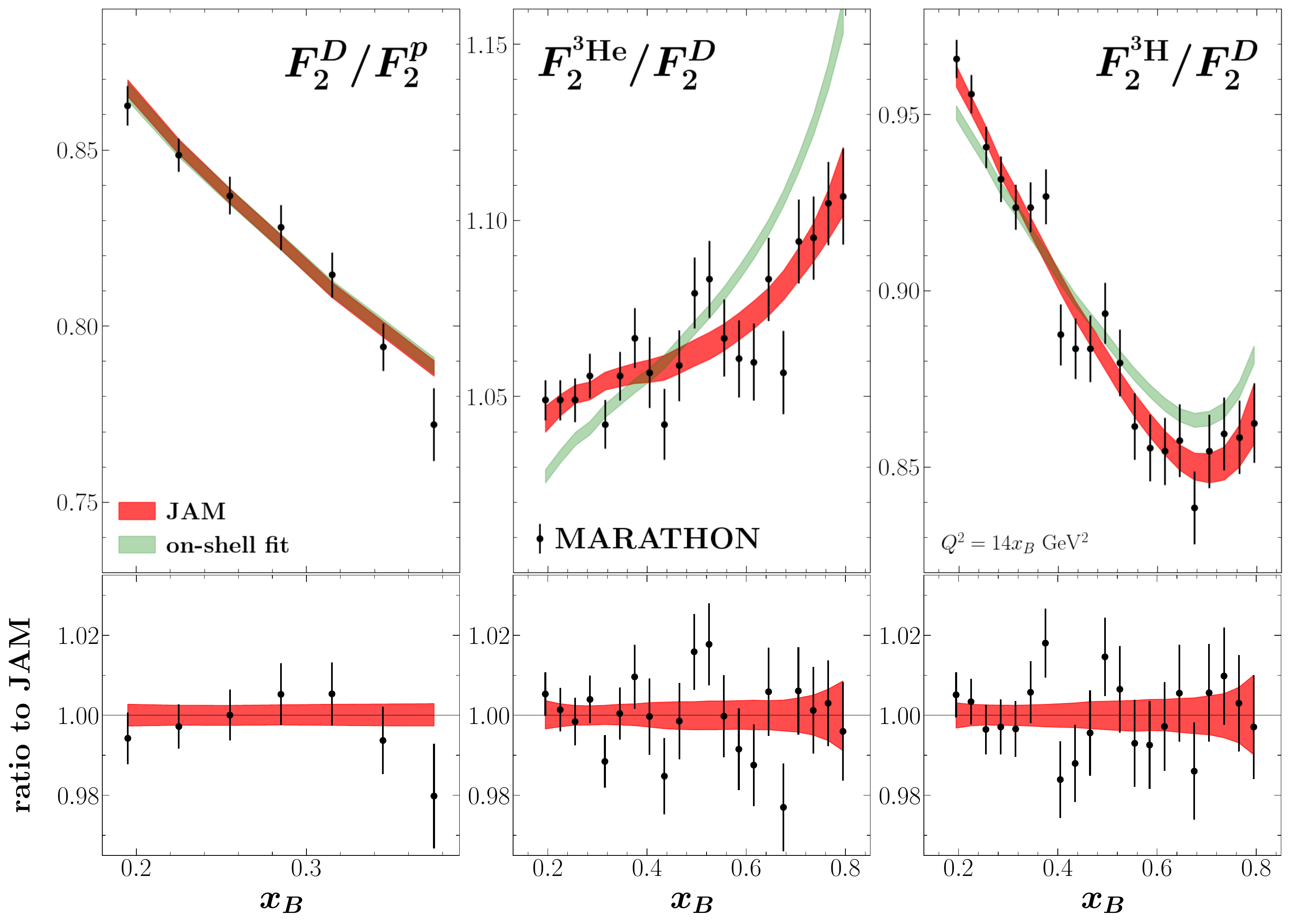}
\caption{Ratios $F_2^D/F_2^p$ (left), $F_2^{\hel}/F_2^{D}$ (middle), and $F_2^{\tri}/F_2^{D}$ (right) from \MAR~\cite{JeffersonLabHallATritium:2021usd, JeffersonLabHallATritium:2024las} (black circles) at the experimental kinematics $Q^2=14 x_B$~GeV$^2$ compared to the full JAM fit (red 68\% CI bands) and to the on-shell fit (green 68\% CI bands).
The bottom panels show the ratios of the results relative to the mean JAM results. The theory curves are normalized by the fitted values 1.017(4), 0.996(6), and 0.991(5) for the $D/p$, $\hel/D$, and $\tri/D$ ratios, respectively.}
\label{f.marathon}
\end{figure}

We stress that the three $A=3$ observables from \MAR ($\hel/\tri$, $\hel/D$, and $\tri/D$) are not independent.
To avoid double-counting of the data, only two of these can be used in the fit.
Since the $\hel/D$ and $\tri/D$ ratio data are the most recent, in practice we choose to include these in the main analysis.
We have checked that selecting other combinations of the data (namely, $\hel/\tri$ and $\hel/D$, or $\hel/\tri$ and $\tri/D$) in the fit does not affect any of the conclusions.


A comparison of the results of the global analysis with the \MAR data is shown in Fig.~\ref{f.marathon}, where the results of the full fit are shown together with an on-shell fit in which all off-shell corrections are set to zero.
As is standard practice in global QCD analyses, the fitted normalizations are used to shift the theory curves, while the experimental data are left unmodified.
Consistent with the observation in Ref.~\cite{Cocuzza:2021rfn}, for the $D/p$ ratio there is little change between the two fits.
For the new $\tri/D$ data, the on-shell fit slightly undershoots the data at lower $\xb$ and slightly overshoots it at larger $\xb$. 
With off-shell corrections included, the data are described very well across all $\xb$.
As expected from the \chired values, the largest impact occurs for the $\hel/D$ ratio, where the on-shell only results severely overshoot the data at large $\xb$.
The inclusion of off-shell corrections resolves this discrepancy.


\begin{table}[b]
\caption{Fitted normalizations from the current JAM analysis compared to the normalizations used in Refs.~\cite{JeffersonLabHallATritium:2021usd, JeffersonLabHallATritium:2024las} from the KP model, along with the $\sigma$ differences between these (with uncertainties summed in quadrature).  The experimental normalization uncertainties from Ref.~\cite{Li:2022fhh} are also shown.}
\begin{tabular}{l c | c c c}
\hhline{=====}
dataset             &
experimental        &
fitted              &
\MAR                &
~~significance~~
\\
                    & 
~~norm. uncertainty \cite{Li:2022fhh}~~   &
~~normalization~~   &
+KP model           &
~~of difference~~
\\
\hline
{\scriptsize MARATHON} {\footnotesize $D$}/$p$ \cite{JeffersonLabHallATritium:2021usd}     
& 0.8\%      & \textbf{1.017(4)}  & --- & ---
\\
{\scriptsize MARATHON} {\footnotesize $\hel$}/$D$ \cite{JeffersonLabHallATritium:2024las} & 1.2\%  & \textbf{0.996(6)}  & 1.021(5) & 3.2$\sigma$
\\
{\scriptsize MARATHON} {\footnotesize $\tri$}/$D$ \cite{JeffersonLabHallATritium:2024las} & 0.8\%  & \textbf{0.991(5)}  & 0.996(5) & 0.7$\sigma$
\\
\hhline{=====}
\end{tabular}
\label{t.norms}
\end{table}

The normalizations for the \MAR structure function ratios for the full off-shell fit are shown in Table~\ref{t.norms}.  
The experimental normalization uncertainties used here were taken from Ref.~\cite{Li:2022fhh}, where they were determined by the uncorrelated (from target to target) target thickness uncertainties combined in quadrature with the uncorrelated beam heating effects.
In contrast, the normalization uncertainties reported in the \MAR publications~\cite{JeffersonLabHallATritium:2021usd, JeffersonLabHallATritium:2024las} were taken from the uncertainty on the size of the normalization shifts required for the derivative $F_2^n/F_2^p$ ratio to agree at $x_B = 0.31$ when extracted from deuterium and $A=3$ ratios using the KP model.
This normalization procedure assumes that the \MAR $F_2^D/F_2^p$ data require no normalization, and that the nuclear effects for all nuclei vanish at $x_B = 0.31$, as in the KP model~\cite{Kulagin:2004ie, Kulagin:2010gd}.
We stress that while the latter behavior has been observed for ratios of structure functions of heavy nuclei to deuterium~\cite{Gomez:1993ri}, there is no direct evidence to suggest that it also holds for $A \leq 3$ nuclei, and, as we show below, enforcing this constraint introduces a significant bias into the analysis.

For the $\hel/D$ and $\tri/D$ ratios, the fitted normalizations are within 1$\sigma$ using the experimental normalization uncertainties~\cite{Li:2022fhh}.
For the $D/p$ ratio, the fitted normalization is 2.2$\sigma$ away from 1, indicating some tension of the \MAR deuteron data with other datasets in the global analysis.
We also compare our normalizations to those derived using the KP model applied to the \MAR data in Refs.~\cite{JeffersonLabHallATritium:2021usd, JeffersonLabHallATritium:2024las}.
For $\hel/D$ we find differences of 3.2$\sigma$, indicating significant tension between our extracted normalization and that obtained using the KP model.
For $\tri/D$, on the other hand, the two normalizations are in better agreement within the uncertainties.

\begin{figure}[t]
\includegraphics[width=0.7\textwidth]{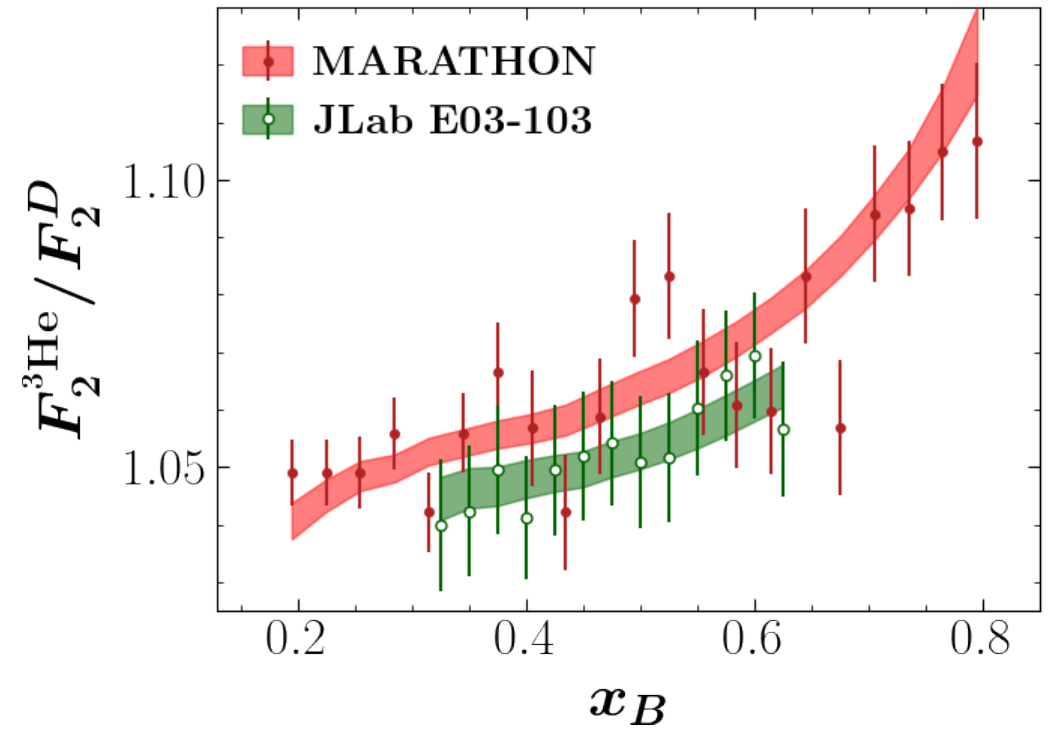}
\caption{Ratio $F_2^{\hel}/F_2^{D}$ from the Jefferson Lab \MAR~\cite{JeffersonLabHallATritium:2024las} (red circles) and Hall~C E03-103~\cite{Seely:2009gt} (green open circles) experiments compared to the JAM fit (colored 68\% CI bands). No normalizations have been applied to any of the data here, while the theory curves have been divided by the normalizations listed in Table~\ref{t.norms}. 
}
\label{f.hd_ratio}
\end{figure}

In Fig.~\ref{f.hd_ratio} we compare the $\hel/D$ data from \MAR and the earlier data from the Jefferson Lab E03-103 experiment~\cite{Seely:2009gt}, and with the fitted results of our global analysis.
No normalizations have been applied to any of the data here, but the theory curves have been divided by the normalizations listed in Table~\ref{t.norms}. 
We find excellent agreement between the experimental results and the fitted values, as well as consistency between the two experiments.
In contrast, the results in Fig.~1 of Ref.~\cite{JeffersonLabHallATritium:2024las} indicate the \MAR $\hel/D$ ratio lying significantly higher, with essentially no overlap with the Jefferson Lab E03-103 data.
This is a direct result of the 2.1\% upward normalization that was applied to the \MAR data in Ref.~\cite{JeffersonLabHallATritium:2024las} in order to match with the KP model result~\cite{Kulagin:2004ie, Kulagin:2010gd}.
From our global analysis, we find that this normalization is not necessary, and in fact introduces a significant bias into the subsequent extraction of the neutron structure function and determination of the nucleon off-shell contributions.


\begin{figure}[t]
\includegraphics[width=0.7\textwidth]{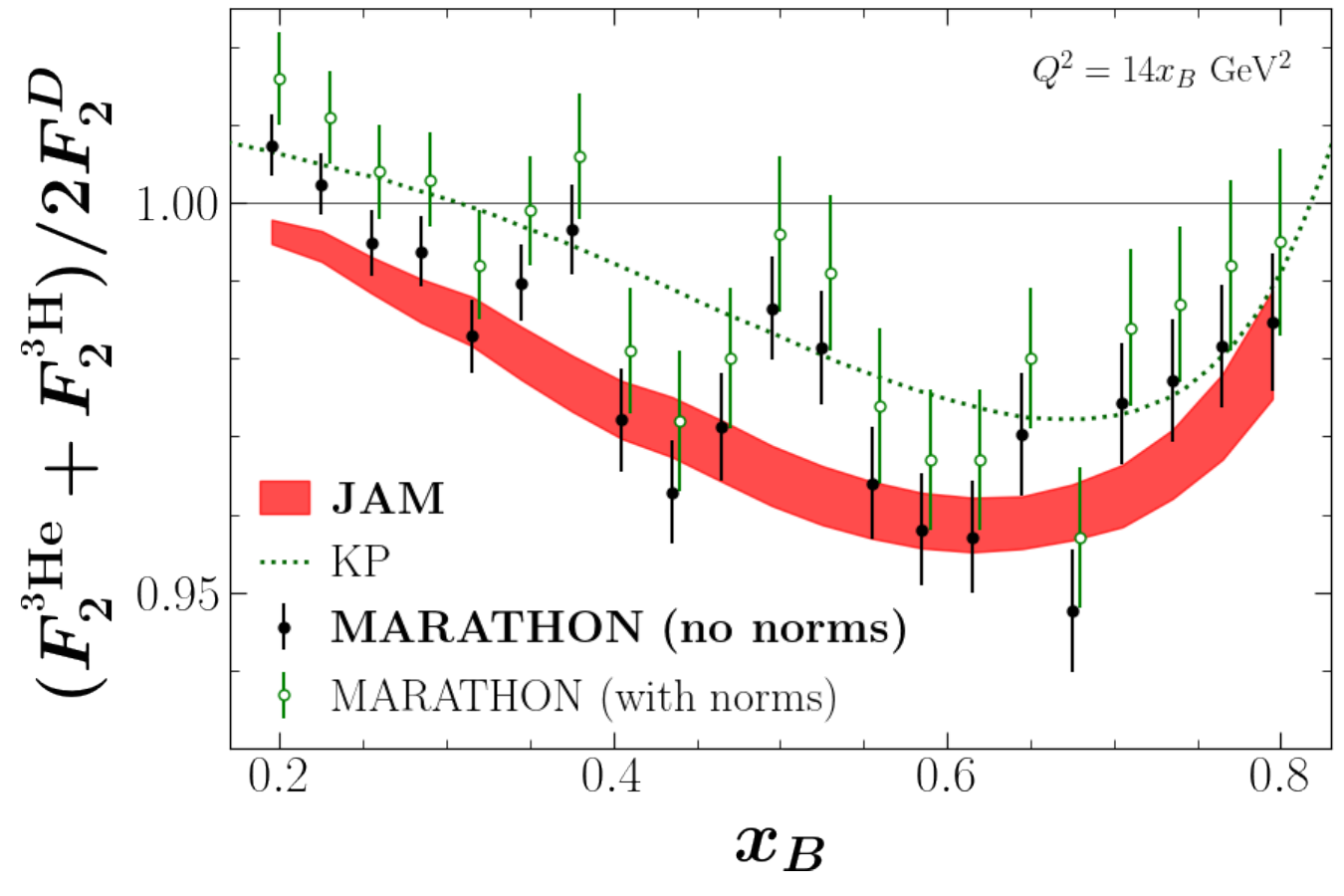}
\caption{Isoscalar nuclear EMC ratio $(F_2^{\hel} + F_2^{\tri})/2F_2^D$ at the \MAR kinematics $Q^2 = 14 \xb$ GeV$^2$.  The JAM result, unmodified by any normalizations, (red 68\% CI band) is compared to the result directly from the \MAR paper \cite{JeffersonLabHallATritium:2024las} (green open circles) and the result where the 1.021 and 0.996 normalizations are removed from $\hel$ and $\tri$, respectively (black circles).  The errors on the latter are computed using the quadrature sum of the uncorrelated errors on the $F_2^{\hel}/F_2^D$ and $F_2^{\tri}/F_2^D$ ratios. The result is also compared to that of the KP model with isoscalar OSE \cite{Kulagin:2010gd, JeffersonLabHallATritium:2024las} (green dotted line).}
\label{f.iso_emc}
\end{figure}

In Fig.~\ref{f.iso_emc} we show the isoscalar EMC ratio $(F_2^{\hel} + F_2^{\tri})/2F_2^D$, as in Fig. 2 of Ref.~\cite{JeffersonLabHallATritium:2024las}.
We compare the JAM result, unmodified by any fitted normalizations, to the \MAR data that are also unmodified by any normalizations.  
The results agree reasonably well, with JAM slightly undershooting it. 

We emphasize, however, that the JAM ratio in Fig.~\ref{f.iso_emc} is a pure theory result unmodified by fitted normalizations, which cannot be cleanly translated from the three \MAR observables ($F_2^D/F_2^p$, $F_2^{\hel}/F_2^D$, and $F_2^{\tri}/F_2^D$) to this new ratio.
The slight underestimate can therefore be explained by the non-negligible normalization on $F_2^D/F_2^p$ of $1.017(4)$, which would decrease $F_2^D$ and thus increase the ratio, which is not applied in Fig.~\ref{f.iso_emc}.
The result from Ref.~\cite{JeffersonLabHallATritium:2024las}, including their normalizations, is also shown and compared to the KP model. 
These also agree well, which is unsurprising given that the normalizations are matcheded by construction.
It interesting to observe that the disagreement between the JAM results and the KP model reflects very closely the difference between the \MAR data without and with the applied normalizations. 
This difference shows clearly the bias that is introduced by applying normalizations to the experimental data that are based on the assumed validity of the KP model for the $A=3$ system.

\section{QCD analysis}
\label{s.QCDanalysis}

Having demonstrated good agreement between the global fit and the \MAR data, in this section we discuss the results of our global QCD analysis, which is based on an ensemble of over 700 replicas, for the on-shell $u$ and $d$ quark PDFs, the isospin dependence of the off-shell corrections, and the systematics of the nuclear EMC ratios for $A=2$ and $A=3$ nuclei.

\subsection{Impact on proton PDFs}
\label{ss.pdfs}

\begin{figure}[t]
\includegraphics[width=\textwidth]{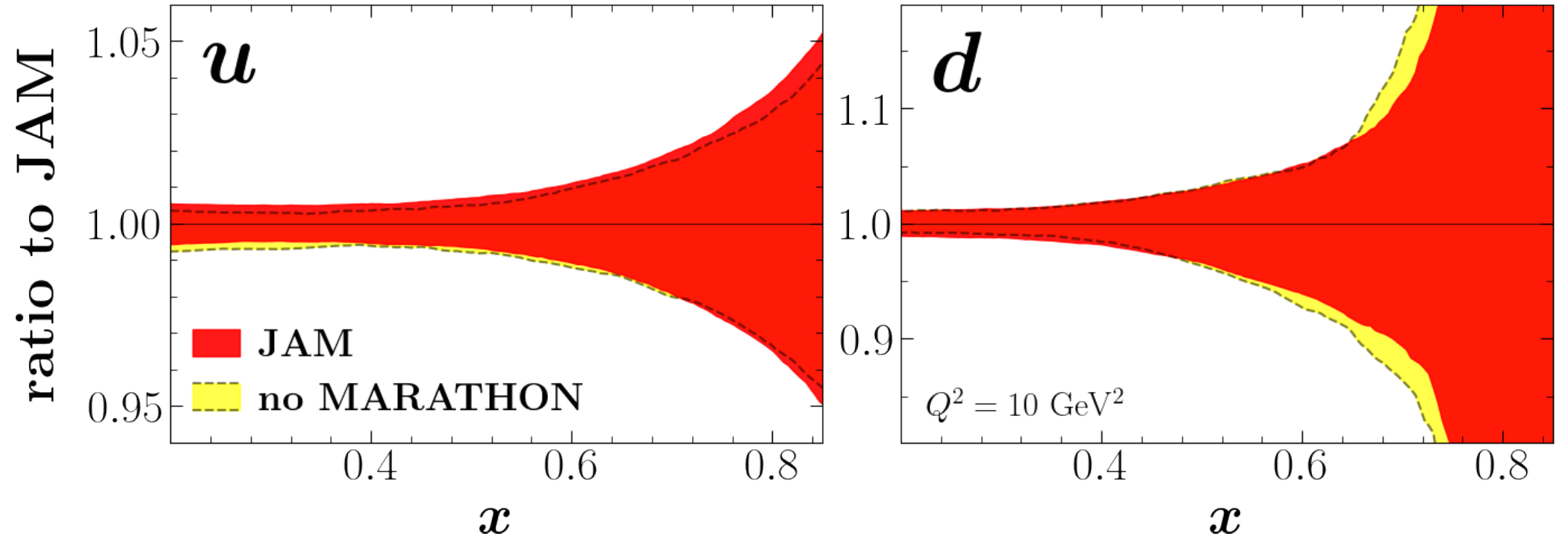}
\vspace*{-1.1cm}
\caption{Ratios of PDFs for $u$ (left) and $d$ (right) quarks relative to the mean JAM result, for the full JAM analysis (red 68\% CI band) and the analysis with no \MAR data included (yellow 68\% CI band) at $Q^2=10$~GeV$^2$.}
\label{f.pdfs}
\end{figure}

The results for the $u$ and $d$ quark PDFs in the proton from our new analysis are shown in Fig.~\ref{f.pdfs} as ratios to the JAM central values, at a scale of $Q^2=10$~GeV$^2$, for analyses with and without the inclusion of the \MAR data.
We observe very little impact of the \MAR data on the PDFs, with almost no change in the central values and similar uncertainties for the two analyses, suggesting that these are already well constrained by other data.
The only visible impact is a small reduction in uncertainties in the $d$ quark PDF at $x \gtrsim 0.6$.
This is consistent with the findings of the previous JAM analysis~\cite{Cocuzza:2021rfn} that included the earlier $\hel/\tri$ \MAR data~\cite{JeffersonLabHallATritium:2021usd}.
As expected, since the $u$ quark PDF is more strongly constrained by proton DIS and other data, the PDF uncertainties are $\lesssim 5\%$ over the entire region of $x \lesssim 0.85$ constrained by data.
For the $d$ quark PDF, the uncertainties are larger, $\approx 20\%$ for $x \approx 0.75$, and increase significantly at higher $x$.

\subsection{Isospin dependence of off-shell corrections}
\label{ss.offshell}

\begin{figure}[t]
\includegraphics[width=0.6\textwidth]{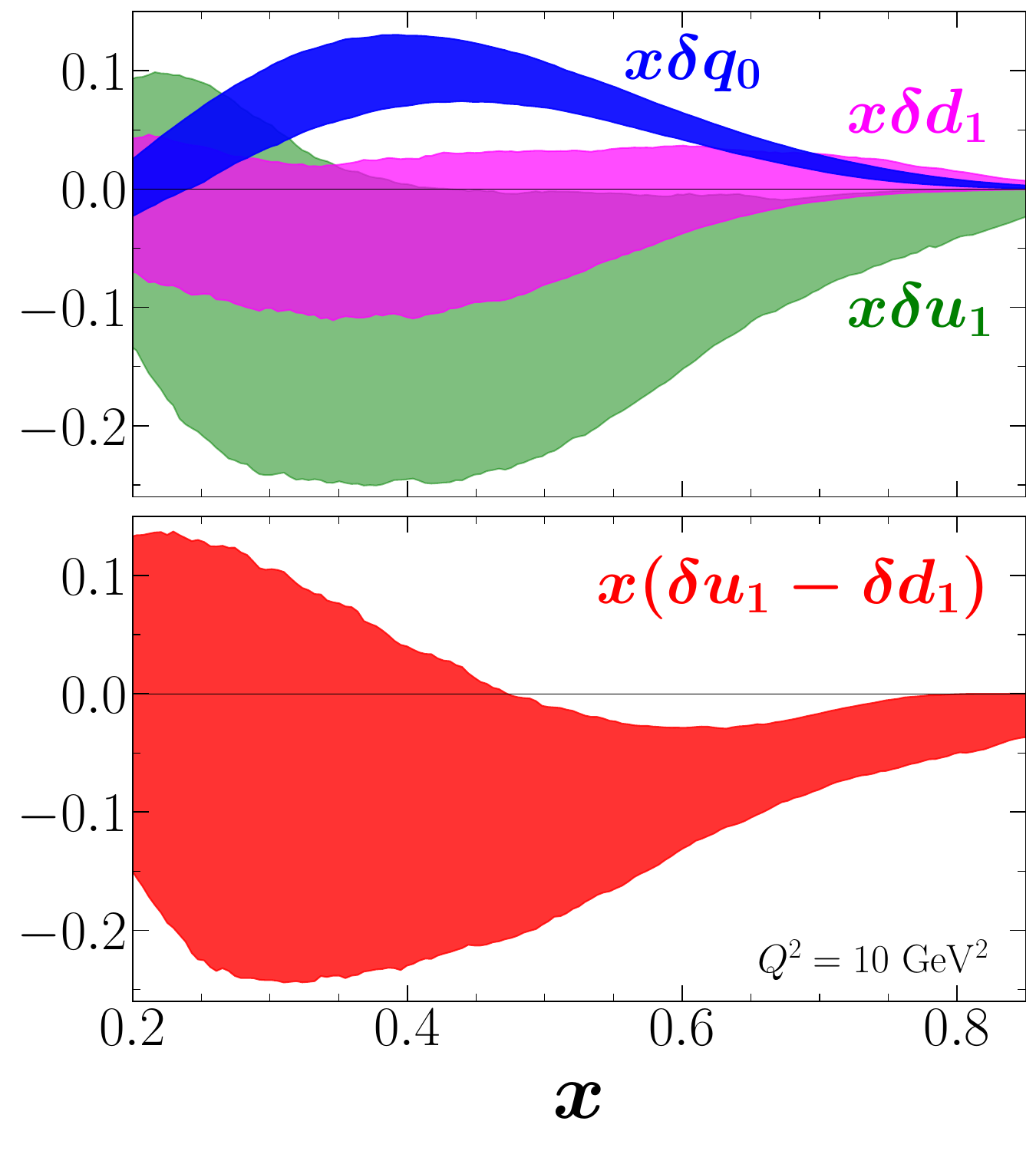}
\caption{Off-shell functions $x \delta q_i$ versus $x$ at $Q^2 = 10$~GeV$^2$:
    (top panel) isoscalar $x \delta q_0 \equiv x \delta u_0 = x \delta d_0$ (blue), isovector $x \delta u_1$ (green), and isovector $x \delta d_1$ (orange) distributions; 
    (bottom panel) difference $x(\delta u_1 - \delta d_1)$ (red).
    The bands indicate 68\% CIs.}
\label{f.offpdfs}
\end{figure}

The off-shell functions $\delta q_i$ from our analysis are shown in Fig.~\ref{f.offpdfs} 
for the isoscalar component $\delta q_0$ and the isovector contributions $\delta u_1$ and $\delta d_1$, as well as for the difference $\delta u_1 - \delta d_1$, at the scale $Q^2 = 10$~GeV$^2$.
Notably, we observe that the isoscalar off-shell correction is nonzero and positive, with $x \delta q_0$ peaking at $x \approx 0.4$.
The isovector contributions have larger uncertainties, with $x \delta d_1$ mostly consistent with zero, but $x \delta u_1$ displaying a negative trend at large~$x$.
This trend is enhanced in the difference $x (\delta u_1 - \delta d_1)$, which indicates a clear negative behavior for $x \gtrsim 0.5$, suggesting the presence of an isovector effect.

\begin{figure}[t]
\includegraphics[width=0.70\textwidth]{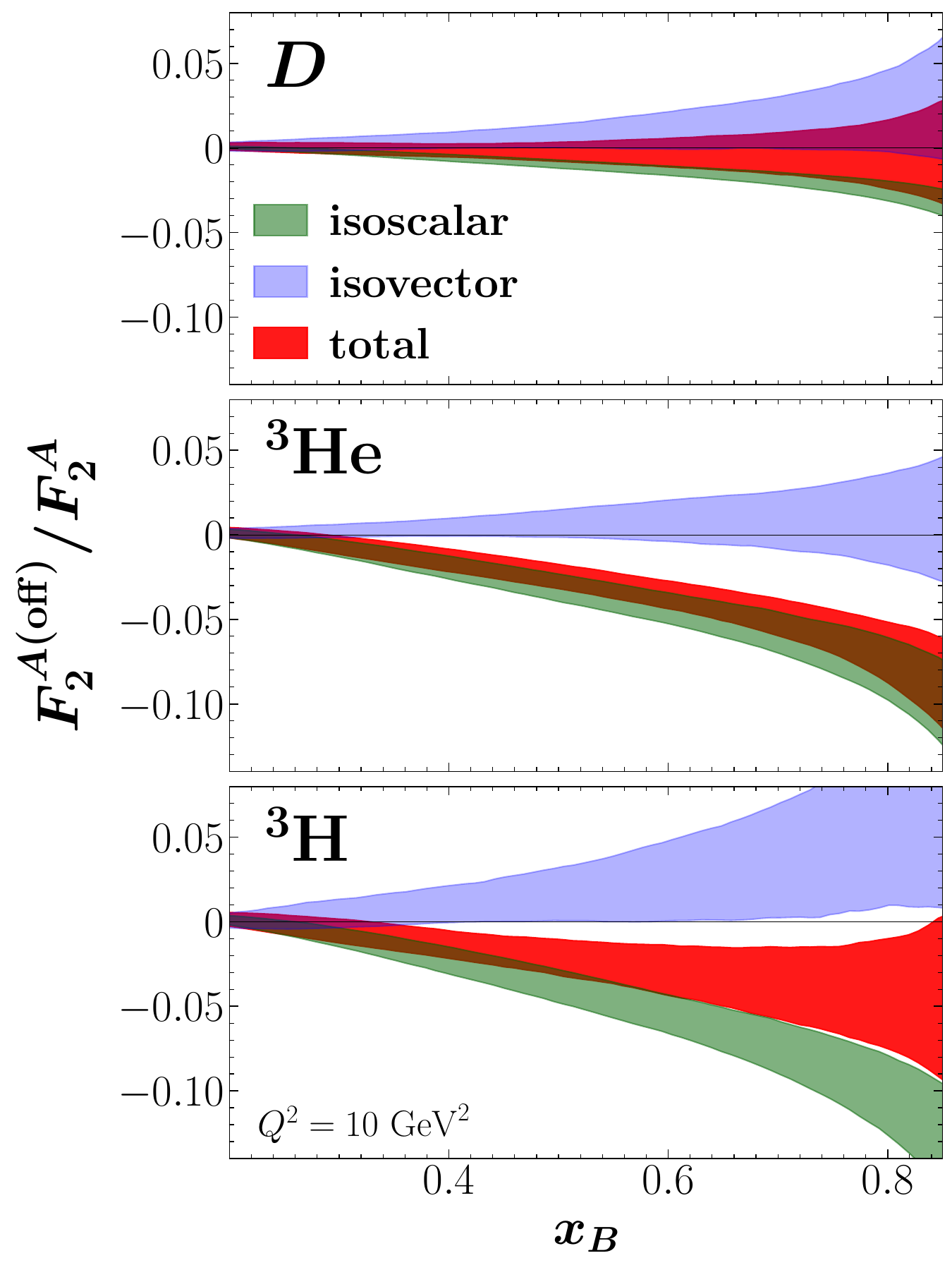}
\caption{Off-shell contributions $F_2^{A({\rm off})}$ to the nuclear structure functions relative to the total $F_2^A$ for $A=D$ (top), $^3$He (middle), and $^3$H (bottom) at $Q^2 = 10$~GeV$^2$ (68\% CI red bands).  The mean values of the isoscalar (68\% CI blue bands) and isovector (68\% CI green bands) contributions are shown separately.}
\label{f.F2A_off}
\end{figure}

The trends in the $\delta q_i$ off-shell functions can be understood by examining their contributions to the nuclear structure functions, as shown in Fig.~\ref{f.F2A_off}.
Here, the off-shell contribution to $F_2^A$ is given by
\begin{eqnarray}
F_2^{A (\rm off)}(x_B,Q^2)
&=& \sum_N \big[ f_{N/A}^{(\rm off)} \otimes \delta F_2^{N/A} \big](x_B,Q^2),
\end{eqnarray}
and is shown as a ratio to the total structure function, at a scale of $Q^2=10$~GeV$^2$, along with the separate contributions from the isoscalar and isovector off-shell components.
The isoscalar contribution is found to be negative for all nuclei considered.
For deuterium it is suppressed by a factor of $\approx 4$ relative to $\hel$ and $\tri$ because of the weaker binding [see Eq.~(\ref{eq.smearing}) below)] and the combinatorical factor 2 in Eq.~(\ref{eq.delta0}), which effectively halves the isoscalar correction for deuterium relative to helium and tritium. 
The effect is a direct consequence of $\delta q_0$ being positive (as the off-shell smearing functions are negative) and the fact that the isoscalar off-shell corrections affect all three nuclei in a similar manner.

The isovector components, on the other hand, are more varied.
They are relatively large and positive for the deuteron and tritium, leading to strong cancellations with the isoscalar component.
This cancellation is almost total for the deuteron, while it leaves a small, negative contribution to $\tri$ at large $\xb$.
For $\hel$, the isovector component is very small, leaving the effect here to be dominated by the large and negative isoscalar component.

These trends can be directly related to those observed in the data versus theory comparisons for the structure function ratios in Fig.~\ref{f.marathon}.
The small effects for the deuteron that largely cancel explain why there is no significant change in the \MAR $D/p$ data, or any of the other deuteron data in this analysis.
For tritium, the isoscalar contribution alone makes the $F_2^{\tri}/F_2^D$ ratio too small at larger $\xb$.
The cancellation with the isovector contribution raises the $F_2^{\tri}/F_2^{D}$ ratio slightly at larger $\xb$.
For helium, the largest contribution by far is from the isoscalar, which significantly reduces $F_2^{\hel}$, and thus the $F_2^{\hel}/F_2^D$ ratio.
When the isovector component is added, the $F_2^{\hel}/F_2^D$ ratio decreases slightly as $\xb$ becomes larger.
While the isovector contribution to $F_2^{\hel}$ is negligible, the slight positive isovector contribution to $F_2^D$ causes the $F_2^{\hel}/F_2^D$ ratio to decrease slightly.

\subsection{Nuclear EMC ratios}
\label{ss.emc}

Since the \MAR experiment only measured ratios of cross sections (structure functions) and not their absolute values, we shall be primarily interested in EMC-type ratios of structure functions of nuclei $A$ to free nucleons.
%
%
In particular, for the $A=2$ and $A=3$ nuclei probed in the \MAR experiment, we define the ``isoscalar'' structure function ratios
\begin{eqnarray}
R_D &=& \frac{F_2^D}{F_2^p + F_2^n},
\end{eqnarray}
for the deuteron, and
\begin{eqnarray}
R_{^3{\rm He}} &=& \frac{F_2^{^3{\rm He}}}{2 F_2^p +   F_2^n}, \qquad\quad
R_{^3{\rm H}}\  =\ \frac{F_2^{^3{\rm H}} }{  F_2^p + 2 F_2^n},
\end{eqnarray}
for $\hel$ and $\tri$, respectively.
Further, taking ratios of the nuclear EMC ratios, we define the helium to tritium super-ratio as
\begin{eqnarray}
{\cal R}_{\hel/\tri} &=& \frac{R_{^3{\rm He}}}{R_{^3{\rm H}}},
\end{eqnarray}
and the analogous super-ratios with respect to the deuteron as
\begin{eqnarray}
{\cal R}_{\hel/D} &=& \frac{R_{^3{\rm He}}}{R_D}, \qquad\quad
{\cal R}_{\tri/D}\ =\ \frac{R_{^3{\rm H}}}{R_D}.
\end{eqnarray}

\begin{figure}[t]
\includegraphics[width=0.80\textwidth]{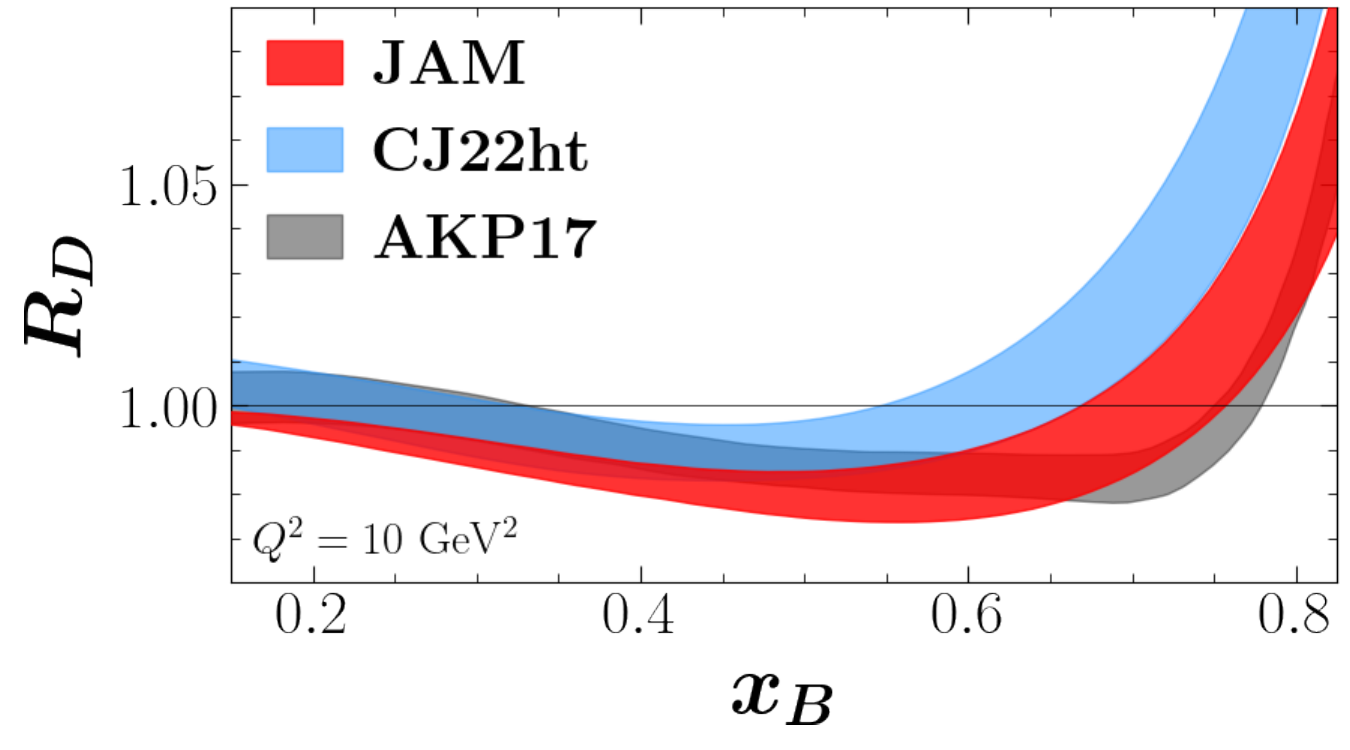}
\caption{Deuteron to free nucleon EMC ratio $R_D$ versus $\xb$ at $Q^2 = 10$~GeV$^2$ from the present JAM analysis (red 68\% CI band), compared with that from the CJ22ht~\cite{Cerutti:2025yji} (blue 90\% confidence band) and AKP17~\cite{Alekhin:2017fpf} (gray 1$\sigma$ band) PDF parametrizations.}
\label{f.emc_D_compare}
\end{figure}

In Fig.~\ref{f.emc_D_compare} we compare the $\xb$ dependence of the deuteron to free nucleon EMC ratio $R_D$ at $Q^2 = 10$~GeV$^2$ from the current JAM analysis with results from the CJ22ht~\cite{Cerutti:2025yji} and AKP17 \cite{Kulagin:2004ie, Kulagin:2010gd, Alekhin:2017fpf} global analyses, the latter of which utilizes constraints from the KP model.
At small $\xb$ the JAM ratio lies slightly below the CJ22ht and AKP17 results, while at larger $\xb \gtrsim 0.6$ the ratio is at the lower end of the CJ22ht band and just above the AKP17 band.
Most notably, there is no indication for a cross of the ratio through unity at $\xb = 0.31$, as assumed in the KP model (see discussion below).
Comparing to the JAM results without the \MAR data, there is a very small reduction in uncertainties on $R_D$ below $\xb \approx 0.4$, approximately matching the upper $\xb$ limit of the \MAR $F_2^D/F_2^p$ data.

\begin{figure}[t]
\includegraphics[width=0.85\textwidth]{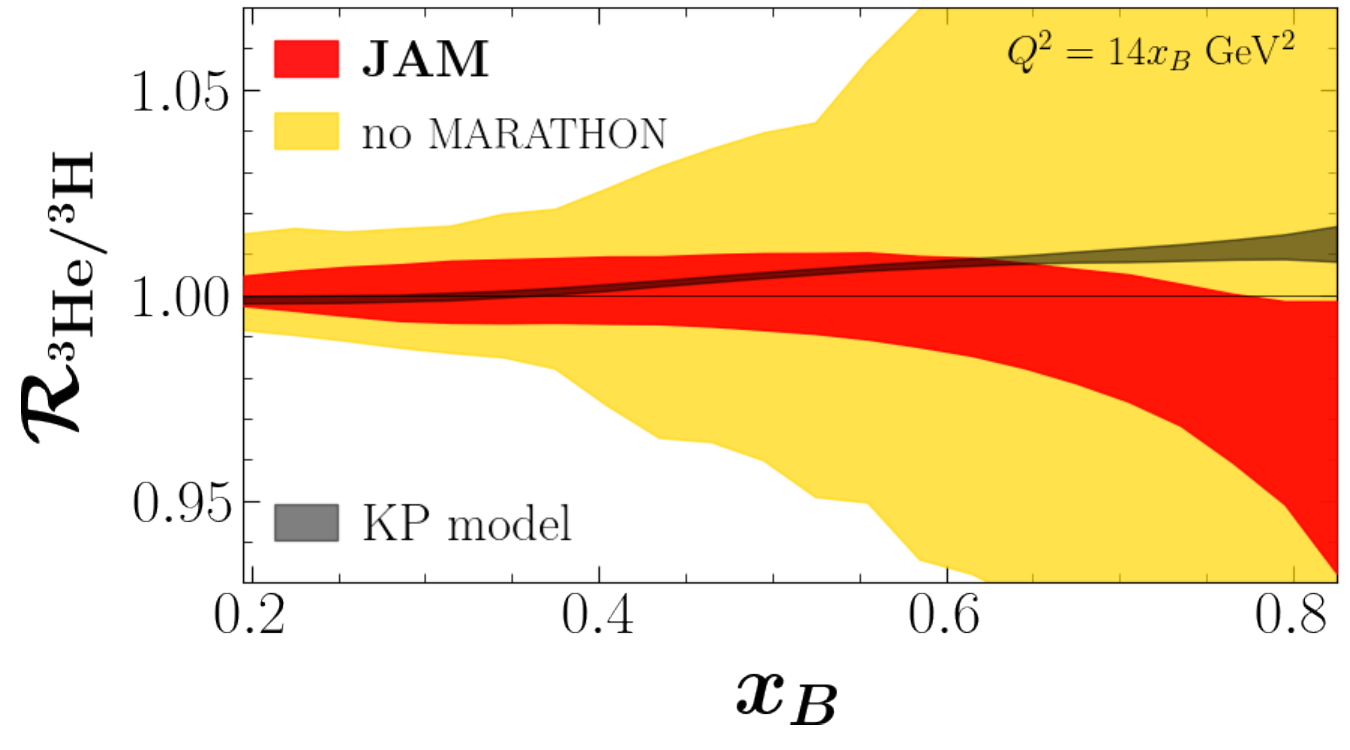}
\caption{Super-ratio ${\cal R}_{\hel/\tri}$ of the $\hel$ to $\tri$ nuclear EMC ratios, as a function of $\xb$ for $Q^2 = 14 \xb$~GeV$^2$, from the present JAM analysis (red 68\% CI band), compared with the ratio without the \MAR data (yellow 68\% CI band), 
and the result from the KP model (gray 1$\sigma$ band)~\cite{Kulagin:2004ie, Kulagin:2010gd} that was used to extract the $F_2^n/F_2^p$ ratio from the \MAR data in Ref.~\cite{JeffersonLabHallATritium:2024las}.}
\label{f.emc_ht_compare}
\end{figure}

For the $A=3$ super-ratio $\mathcal{R}_{\hel/\tri}$, shown in Fig.~\ref{f.emc_ht_compare} at the experimental kinematics $Q^2 = 14 \xb$~GeV$^2$, at intermediate $\xb$ values our result largely overlaps with the KP model that was used to extract the $F_2^n/F_2^p$ ratio in Ref.~\cite{JeffersonLabHallATritium:2021usd}, but deviates at larger values of $\xb$.
In particular, our result tends towards values lower than 1, whereas the KP model remains above unity up to $\xb \approx 0.8$, similar to what was observed in earlier calculations of the super-ratio that assumed flavor independent off-shell corrections in $A=3$ nuclei~\cite{Afnan:2000uh, Afnan:2003vh, Pace:2001cm, Sargsian:2001gu}.
Our fitted result is consistent with the findings in the previous JAM analysis~\cite{Cocuzza:2021rfn}, except now the super-ratio has larger errors because of the increased flexibility of the off-shell model.
Comparing to the result without the \MAR data, one sees a significant reduction in uncertainties when \MAR data are included.
This is expected, given that prior to the inclusion of the \MAR data there were no direct constraints on $F_2^{\tri}$ at all, and only the Jefferson Lab E03-103 $\hel/D$ data~\cite{Seely:2009gt} existed to constrain $F_2^{\hel}$.

\begin{figure}[t]
\includegraphics[width=0.7\textwidth]{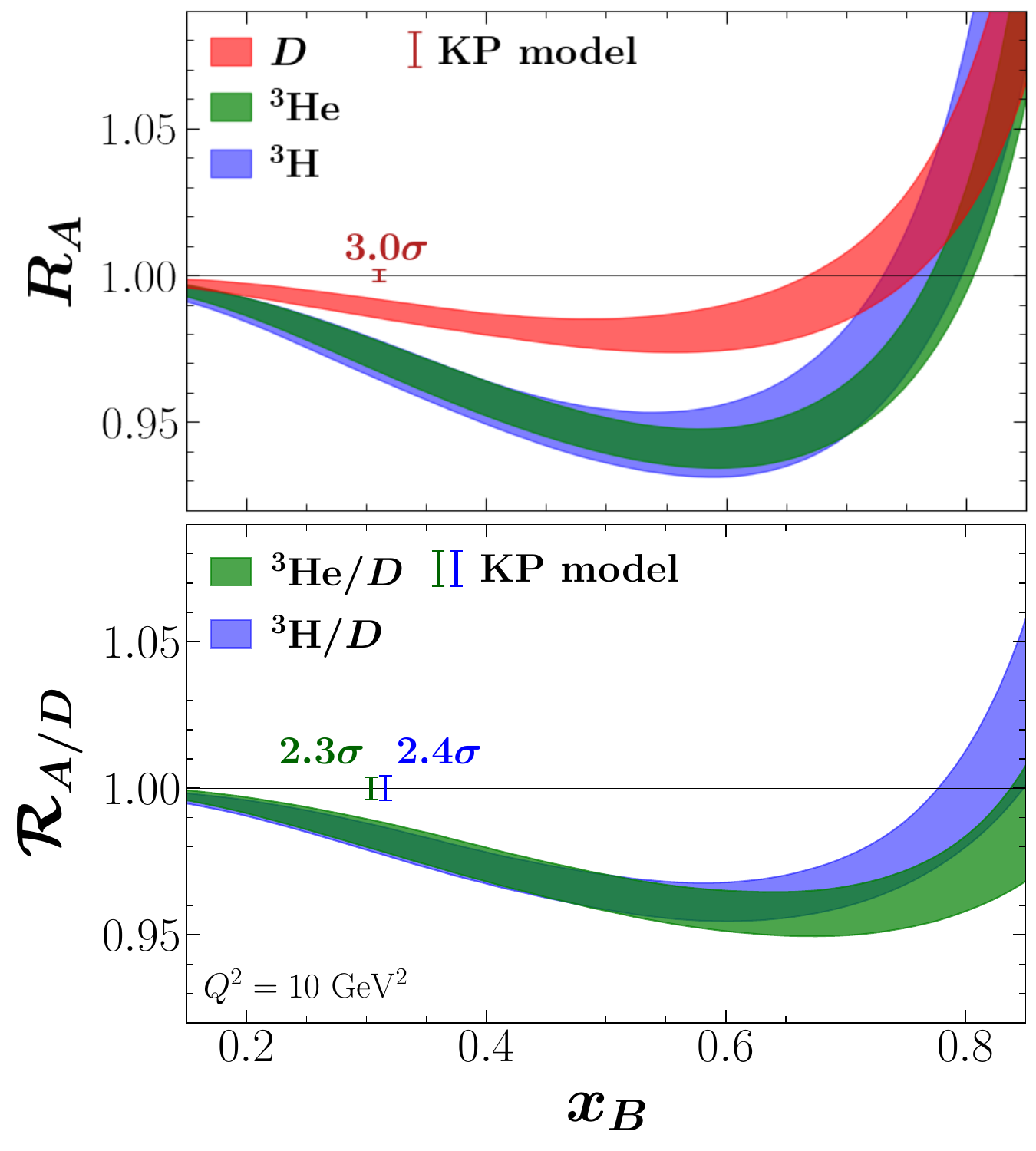}
\caption{
(Top panel) Nuclear EMC ratios $R_A$ for $A=D$ (red band), $\hel$ (green band), and $\tri$ (blue band) for the full JAM analysis. The $\sigma$ difference between the JAM band for $D$ at $x_B = 0.31$ and the KP result (denoted by the red error bar) is calculated using the quadrature sum of the errors.
(Bottom panel) Super-ratios $\mathcal{R}_{A/D}$ for $\hel/D$ (green band) and $\tri/D$ (blue band), with $\sigma$ differences with the KP points for $\hel/D$ (green error bar) and $\tri/D$ (blue error bar) at $x_B = 0.31$ indicated. All ratios are given at $Q^2 = 10$~GeV$^2$, and all bands represent 68\% CIs.}
\label{f.emc_rat}
\end{figure}

The EMC ratios $R_A$ for $A=D$, $\hel$, and $\tri$, as well as the super-ratios ${\cal R}_{A/D}$ for $\hel/D$ and $\hel/D$, are shown in Fig.~\ref{f.emc_rat} for the full JAM analysis, along with the on-shell only results and the predictions from the KP model.
The KP model gives unity for each of the ratios $R_D$, ${\cal R}_{\hel/D}$, and ${\cal R}_{\tri/D}$ at $\xb = 0.31$, with uncertainties of 0.20\%, 0.38\%, and 0.42\%, respectively~\cite{JeffersonLabHallATritium:2024las}.
Comparing our uncertainty bands at $\xb = 0.31$ to these points, we find differences of $3.0\sigma$, $2.3\sigma$, and $2.4\sigma$, respectively.
Our result for $R_D$ is $\approx 1\%$ below unity, while the results for $R_{\hel}$ and $R_{\tri}$ are a further $1\%-2\%$ below $R_D$, leading to super-ratios ${\cal R}_{\hel/D}$ and ${\cal R}_{\tri/D}$ that are $\approx 1\%-2\%$ below 1 at $\xb = 0.31$.
Note that the effects for helium and tritium being similar and significantly larger than that for deuterium reflects the larger binding energies and smearing effects for the $A=3$ system.
For the Paris~\cite{Lacombe:1981eg} deuteron wave function and the KPSV~\cite{Kievsky:1996gz} $^3$He spectral function, for example, we find
\begin{eqnarray}
\langle f_{p/D}^{\rm (off)} \rangle \approx -4.3\%, \qquad
%
%
\langle f_{p/{^3}{\rm He}}^{\rm (off)} \rangle \approx -6.8\%,\qquad
%
%
\langle f_{p/{^3}{\rm H}}^{\rm (off)} \rangle \approx -9.5\%. 
%
%
\label{eq.smearing}
\end{eqnarray}

\begin{figure}[t]
\includegraphics[width=1.01\textwidth]{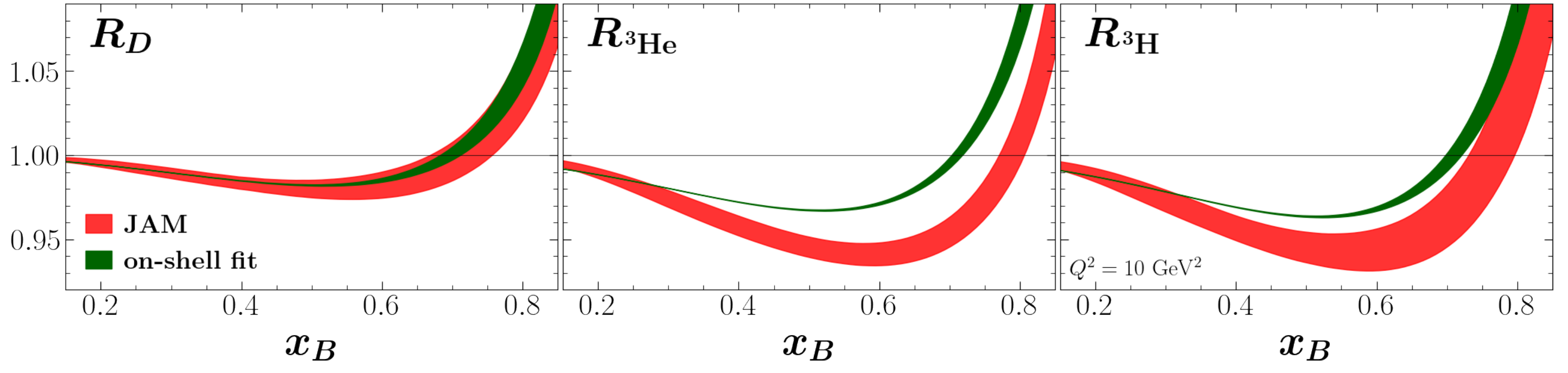}
\caption{Nuclear EMC ratios $R_A$ for $A=D$ (left panel), $\hel$ (middle panel), and $\tri$ (right panel) for the full JAM analysis (red 68\% CI bands) and for the on-shell only fit (green 68\% CI bands) at $Q^2 = 10$~GeV$^2$.}
\label{f.emc_onshell}
\end{figure}

Comparing the full JAM results with the mean on-shell results in Fig.~\ref{f.emc_onshell}, the uncertainties for $R_D$ increase significantly when off-shell corrections are included, while the central value remains largely unchanged.
For $R_{\hel}$ and $R_{\tri}$, the uncertainties also increase while the curves move downward significantly.
These observations are consistent with those seen in Fig.~\ref{f.marathon} when off-shell corrections were included.

In a recent analysis, Alekhin, Kulagin and Petti~\cite{Alekhin:2024dqu} claimed that the results of the previous JAM analysis~\cite{Cocuzza:2021rfn} of the $\hel/\tri$ \MAR data may be biased by the choice of multiplicative HT parametrization and by correlations between off-shell PDF parameters.
While the previous JAM analysis~\cite{Cocuzza:2021rfn} used a multiplicative ansatz for the HT correction, an additive parametrization was also tested and produced no qualitative changes in any of the results.
In the current analysis, we use additive HT corrections, finding results to be consistent with a multiplicative ansatz, and again in disagreement with the KP model.

The authors of Ref.~\cite{Alekhin:2024dqu} also claim that there is an increase in correlations between parameters introduced when including nuclear dependence in the off-shell functions for individual quark flavors, as in the current and previous analyses.
We do find that the central values of the nuclear EMC ratios, as well as their uncertainties, are significantly influenced by the modeling of the off-shell functions, with added flexibility increasing the uncertainties on the ratios.
This can already be seen in Fig.~\ref{f.emc_onshell} when comparing the on-shell and off-shell results.
To obtain the most faithful representation of the uncertainties on the nuclear EMC ratios, it is important to use an off-shell model that is sufficiently flexible given the physical constraints and currently available experimental data, as done in this analysis.
In fact, choosing an insufficiently flexible model, as in Ref.~\cite{Alekhin:2024dqu}, would lead to a biased result with underestimated uncertainties on the nuclear EMC ratios. \\

\section{Conclusions}
\label{s.outlook}

We have performed a global QCD analysis of the world's unpolarized proton, deuteron and $A=3$ data within the JAM Bayesian Monte Carlo framework, including for the first time the latest results from the \MAR experiment at Jefferson Lab on the $\hel/D$ and $\tri/D$ structure functions ratios~\cite{JeffersonLabHallATritium:2024las}.
Most notably, we find that the $\hel/D$ ratio data cannot be described without the inclusion of nucleon off-shell corrections, without which we obtain a \chired of 5.10 for these data, while inclusion of off-shell corrections greatly improves the \chired to 1.04.
These data clearly indicate the necessity of off-shell contributions to describe all of the world DIS data on $A \leq 3$ nuclei.
A smaller improvement is also observed in the description of the $\tri/D$ data with the inclusion of off-shell corrections.
Within our theoretical framework, which allows for both isoscalar and isovector off-shell corrections, we find that the combination of $\hel/D$ and $\tri/D$ data suggest a sizable isoscalar contribution, with a slight preference for a nonzero isovector off-shell effect, as indicated by a nonzero $\delta u_1 - \delta d_1$ distribution.

Our results differ significantly from those in the KP model, which is assumed in the analyses of the \MAR data in Refs.~\cite{JeffersonLabHallATritium:2021usd, JeffersonLabHallATritium:2024las}.
Our fitted normalization for $\hel/D$ of $0.996(6)$ is consistent with 1, and differs by $3.2\sigma$ from the KP model result of $1.021(5)$.
For the nuclear EMC ratios, we find a different trend for ${\cal R}_{ht}$ at large $\xb$ compared to that in the KP model, consistent with the findings in the previous JAM analysis~\cite{Cocuzza:2021cbi}.
The KP model also enforces each of the ratios $R_D$, ${\cal R}_{hd}$, and ${\cal R}_{td}$ to be unity 1 at $\xb = 0.31$, while our global analysis disagrees with those results by $3.0\sigma$, $2.3\sigma$, and $2.4\sigma$, respectively.
The fact that our analysis describes the \MAR data well (along with all other experimental data), and leads to different results for quantities such as the nuclear EMC ratios, suggests that the additional normalizations imposed on the \MAR data to be consistent with the KP model needlessly bias the conclusions of the experimental analysis \cite{JeffersonLabHallATritium:2021usd, JeffersonLabHallATritium:2024las}.

Of course, our conclusions must be viewed within the theoretical framework on which our analysis is based.
This includes the basic assumption that DIS from nuclei proceeds through incoherent scattering from individual (off-shell) nucleons, and that multiple scattering from more than one nucleon or scattering from non-nucleonic constituents of the nucleus are relatively rare events, which for light nuclei and at $\xb \gg 0$ should be a very good approximation~\cite{Badelek:1991qa,
Melnitchouk:1992eu, Melnitchouk:1993vc, Badelek:1994qg, Melnitchouk:1995am, Piller:1995kh, Edelmann:1997ik, Bissey:2001cw, Piller:1999wx}.
A further underlying assumption in our analysis is that of charge symmetry for both the on-shell and off-shell PDFs: while the former could be tested with future data on $\pi^\pm$ leptoproduction or charge-changing neutrino interactions from isoscalar targets~\cite{Londergan:2009kj, Hobbs:2011vy}, constraining charge symmetry violating effects in the off-shell functions is unlikely to be realized in the near future.
We also do not consider off-shell corrections to the HT contributions, which would not be possible to determine from existing data.
Finally, while we acknowledge potential bias from the choice of a particular functional form for our PDF parameterization, the use of Monte Carlo sampling in generating many hundreds of replicas for the PDFs in practice reduces any bias from adopting this particular functional form.

On the experimental front, in the near future constraints on neutron structure and the $d/u$ PDF ratio at large $x$ may come from the BONuS12 experiment at Jefferson Lab, which tags spectator protons in semi-inclusive DIS from the deuteron~\cite{Albayrak:2024vcy}, as well as parity-violating DIS measurements on protons through the $\gamma Z$ interference process~\cite{Hobbs:2008mm, Whitehill:2026swa}.
Future data on inclusive and semi-inclusive DIS from $A=3$ nuclei may also provide further information about the isospin dependence of the nucleon structure modifications in nuclei, and shed light on the elusive dynamical origin of the EMC effect.

\clearpage
\begin{acknowledgments}
We thank C.~E.~Keppel and A.~Metz for helpful discussions.
This work was supported by the US Department of Energy Contract No.~DE-AC05-06OR23177, under which Jefferson Science Associates, LLC operates Jefferson Lab, and the National Science Foundation under grant number PHY-1516088.
The work of A.W.T. was supported by Adelaide University and by the Australian Research Council through the Discovery Project DP230101791 and the ARC Centre of Excellence for Dark Matter Particle Physics CE200100008.
The work of N.S. was supported by the DOE, Office of Science, Office of Nuclear Physics in the Early Career Program.
\end{acknowledgments}

\bibliography{cc.bib}

\end{document}